\documentstyle[12pt]{article} 
\input psfig.sty

\def\Title#1{\begin{center} {\Large #1 } \end{center}}
\def\Author#1{\begin{center}{ \sc #1} \end{center}}
\def\Address#1{\begin{center}{ \it #1} \end{center}}

\def\doeack{\footnote{Work supported by the Department of Energy,
                     contract DE--AC03--76SF00515.}}
\def\SLAC{Stanford Linear Accelerator Center\\
    Stanford University, Stanford, California 94309 USA}

\newenvironment{Abstract}{\begin{quotation} \begin{center}
                       ABSTRACT
     \end{center}\bigskip  }{\end{quotation}}
\def\beq{\begin{equation}}
\def\eeq#1{\label{#1}\end{equation}}
\def\eeqn{\end{equation}}
\def\beqa{\begin{eqnarray}}
\def\eeqa#1{\label{#1}\end{eqnarray}}
\def\eeqan{\end{eqnarray}}

\def\Acknowledgements{\bigskip  \bigskip \begin{center} \begin{large}
             \bf ACKNOWLEDGEMENTS \end{large}\end{center}}

\def\Re{{\cal R \mskip-4mu \lower.1ex \hbox{\it e}\,}}
\def\Im{{\cal I \mskip-5mu \lower.1ex \hbox{\it m}\,}}
\def\nn{\noindent}
\def\ie{{\it i.e.}}
\def\eg{{\it e.g.}}

\def\etal{{\it et al.}}

\def\sub#1{_{\lower.25ex\hbox{$\scriptstyle#1$}}}
\def\sul#1{_{\kern-.1em#1}}
\def\sll#1{_{\kern-.2em#1}}  
\def\sbl#1{_{\kern-.1em\lower.25ex\hbox{$\scriptstyle#1$}}}
\def\ssb#1{_{\lower.25ex\hbox{$\scriptscriptstyle#1$}}}
\def\sbb#1{_{\lower.4ex\hbox{$\scriptstyle#1$}}}

\def\to{\rightarrow}
\def\dk{\ifmmode \Delta\kappa\else $\Delta\kappa$\fi}
\def\sigt{\ifmmode \tilde\sigma\else $\tilde\sigma$\fi}
\def\mh{\ifmmode m\sbl H \else $m\sbl H$\fi}
\def\mch{\ifmmode m_{H^\pm} \else $m_{H^\pm}$\fi}
\def\mt{\ifmmode m_t\else $m_t$\fi}
\def\mc{\ifmmode m_c\else $m_c$\fi}
\def\mz{\ifmmode M_Z\else $M_Z$\fi}
\def\mw{\ifmmode M_W\else $M_W$\fi}
\def\mws{\ifmmode M_W^2 \else $M_W^2$\fi}
\def\mhs{\ifmmode m_H^2 \else $m_H^2$\fi}   
\def\mzs{\ifmmode M_Z^2 \else $M_Z^2$\fi}
\def\mts{\ifmmode m_t^2 \else $m_t^2$\fi}
\def\mcs{\ifmmode m_c^2 \else $m_c^2$\fi}
\def\mchs{\ifmmode m_{H^\pm}^2 \else $m_{H^\pm}^2$\fi}
\def\ztwo{\ifmmode Z_2\else $Z_2$\fi}
\def\zone{\ifmmode Z_1\else $Z_1$\fi}
\def\mtwo{\ifmmode M_2\else $M_2$\fi}
\def\mone{\ifmmode M_1\else $M_1$\fi}
\def\tb{\ifmmode \tan\beta \else $\tan\beta$\fi}
\def\xw{\ifmmode x\sub w\else $x\sub w$\fi}
\def\ch{\ifmmode H^\pm \else $H^\pm$\fi}
\def\lum{\ifmmode {\cal L}\else ${\cal L}$\fi}
\def\inpb{\ifmmode {\rm pb}^{-1}\else ${\rm pb}^{-1}$\fi}
\def\infb{\ifmmode {\rm fb}^{-1}\else ${\rm fb}^{-1}$\fi}
\def\epem{\ifmmode e^+e^-\else $e^+e^-$\fi}
\def\ppb{\ifmmode \bar pp\else $\bar pp$\fi}

\def\bsg{\ifmmode b\rightarrow s\gamma \else $b\rightarrow s\gamma$\fi}
 
\newskip\zatskip \zatskip=0pt plus0pt minus0pt
\def\matth{\mathsurround=0pt}

\def\atversim#1#2{\lower0.7ex\vbox{\baselineskip\zatskip\lineskip\zatskip
  \lineskiplimit 0pt\ialign{$\matth#1\hfil##\hfil$\crcr#2\crcr\sim\crcr}}}

\begin{document}
\rightline{\vbox{\halign{&#\hfil\cr
&SLAC-PUB-7298\cr
&September 1996\cr}}}
\vspace{0.8in} 
\Title{The Polarization Asymmetry in $\gamma e$ Collisions at the NLC and 
Triple Gauge Boson Couplings} 
\bigskip
\Author{Thomas G. Rizzo\doeack}
\Address{\SLAC}
\bigskip
\begin{Abstract}
 
The capability of the NLC in the $\gamma e$ collider mode to probe the 
CP-conserving $\gamma WW$ and $\gamma ZZ$ anomalous couplings through the 
use of the polarization asymmetry is examined. When combined with other 
measurements, very strong constraints on both varieties of anomalous couplings 
can be obtained. We show that these bounds are complementary to those that can 
be extracted from data taken at the LHC.

\end{Abstract}
\bigskip
\vskip1.0in
\begin{center}
To appear in the {\it Proceedings of the 1996 DPF/DPB Summer Study on New
 Directions for High Energy Physics-Snowmass96}, Snowmass, CO, 
25 June-12 July, 1996. 
\end{center}
%
\bigskip
\def\thefootnote{\fnsymbol{footnote}}
\setcounter{footnote}{0}
\newpage
\section{Introduction}

As confirmed by the discovery of the top quark, the Standard Model(SM) 
continues to do an excellent job at describing essentially all existing 
data{\cite {tev,blondel}}. In addition to unravelling the source of symmetry 
breaking, one of the most crucial remaining set of tests of the 
structure of the SM will occur at future colliders when precision measurements 
of the various triple gauge boson vertices(TGVs) become 
available{\cite {rev}}. Such analyses are in their infancy today at both the 
Tevatron and LEP. If new physics arises at or near the TeV scale, 
then on rather general grounds one expects that the deviation of the 
TGVs from their canonical SM values, \ie, the anomalous 
couplings, to be {\it at most} ${\cal O}(10^{-3}-10^{-2})$ with the smaller 
end of this range of values being the most likely. To get to 
this level of precision and beyond, for all of the TGVs, a number of 
different yet complementary reactions 
need to be studied using as wide a variety of observables as possible. 

In the present analysis we concentrate on the 
CP-conserving $\gamma WW$ and $\gamma ZZ$ anomalous couplings that can be 
probed in the reactions $\gamma e \to W\nu ,Ze$ at the NLC using polarized 
electrons and polarized backscattered laser photons{\cite {old}}. In the 
$\gamma WW$ case, the anomalous 
couplings modify the magnitude and structure of the already existing SM tree 
level vertex. No corresponding tree level $\gamma ZZ$ vertex exists in 
the SM, although it will appear at the one-loop level. One immediate 
advantage of the $\gamma e\to W\nu$ process over, \eg, 
$e^+e^-\to W^+W^-$ is that the $\gamma WW$ vertex can be trivially isolated 
from the corresponding ones for the $ZWW$ vertex, thus allowing us to probe 
this particular vertex in a model-independent fashion. In addition, the 
$\gamma e\to W\nu$ process probes the TGVs for on-shell photons whereas 
$e^+e^-\to W^+W^-$ probes the couplings at $q^2 \geq 4M_W^2$. 
To set the notation for 
what follows, we recall that the $CP-$conserving $\gamma WW$ and 
$\gamma ZZ$ anomalous couplings are traditionally 
denoted by $\Delta \kappa$, $\lambda$  and $h_{3,4}^0${\cite {rev}}, 
respectively. We will assume that the $\gamma WW$ and $\gamma ZZ$ 
anomalous couplings are unrelated; the full details of our analysis can 
be found in Ref.{\cite {old}}.

\section{Analysis}

The use of both polarized electron and photon beams available at the NLC 
allows one to construct a polarization asymmetry, $A_{pol}$. As we will see, 
this asymmetry provides a new handle on possibly anomalous TGVs of both the 
$W$ and $Z$. In general the $\gamma e \to W\nu ,Ze$ 
(differential or total) cross sections can be written schematically 
as 
\begin{equation}
\sigma=(1+A_0P)\sigma_{un}+\xi(P+A_0)\sigma_{pol} \,,
\end{equation}
where $P$ is the electron's polarization(which we take to be $>0$ for 
left-handed beam polarization), 
$-1\leq \xi \leq 1$ is the Stoke's parameter for the circularly polarized 
photon, and $A_0$ describes the electron's coupling to the relevant gauge 
boson[$A_0=2va/(v^2+a^2)=1$ for $W$'s and $\simeq 0.150$ for $Z$'s, with $v,a$ 
being the vector and axial-vector coupling of the electron]. 
$\sigma_{pol}(\sigma_{un})$ 
represents the polarization (in)dependent contribution to the cross section, 
both of which are functions of only a single dimensionless variable at the 
tree level 
after angular integration, \ie, $x=y^2=s_{\gamma e}/M_{W,Z}^2$,  
where $\sqrt {s_{\gamma e}}$ is the $\gamma -e$ center of mass energy. 
Taking the ratio of the $\xi$-dependent to $\xi$-independent terms in the 
expression for $\sigma$ above gives us the asymmetry $A_{pol}$.

\vspace*{-0.5cm}
\nn
\begin{figure}[htbp]
\centerline{
\psfig{figure=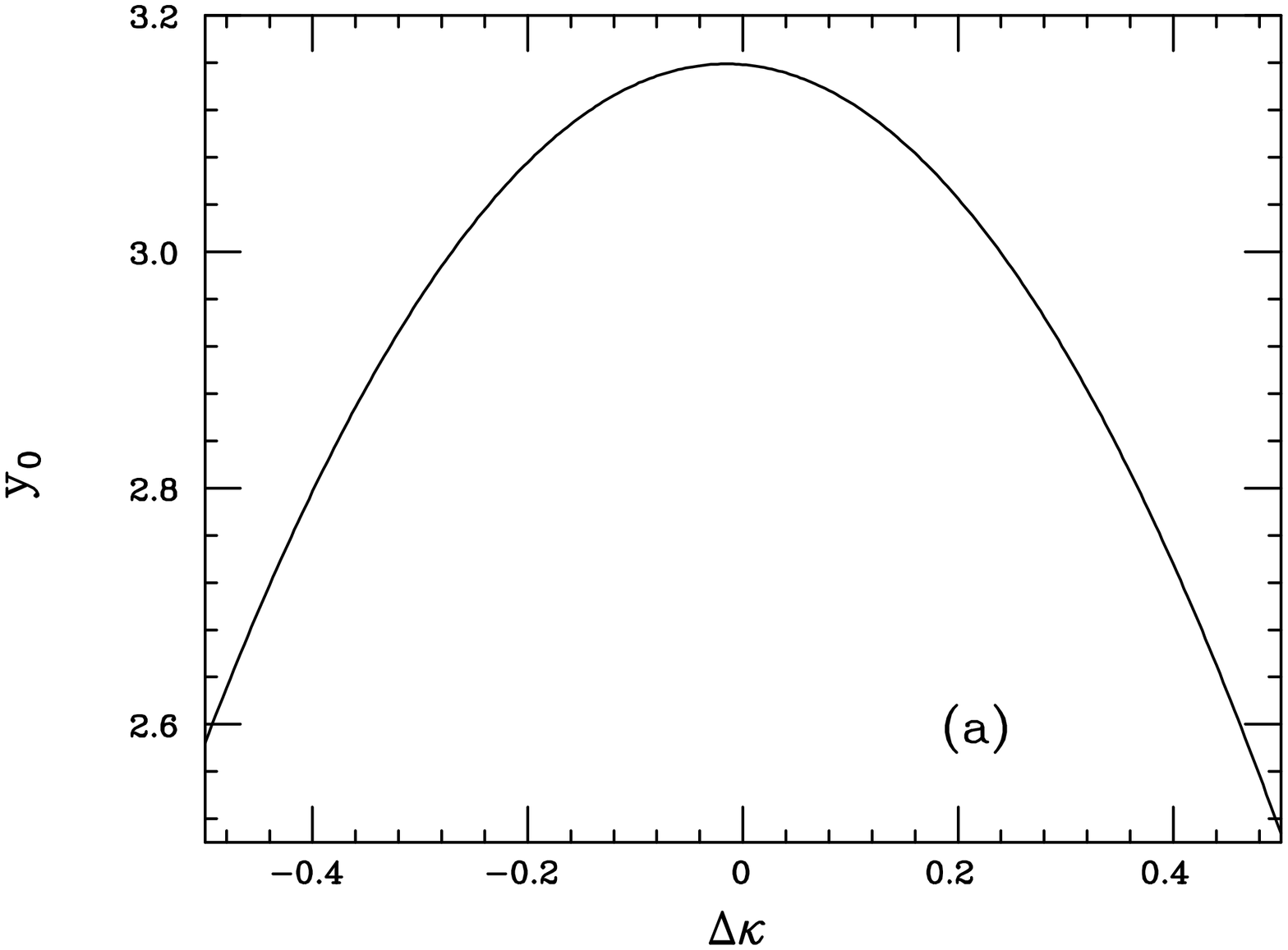,height=9.1cm,width=9.1cm,angle=90}
\hspace*{-5mm}
\psfig{figure=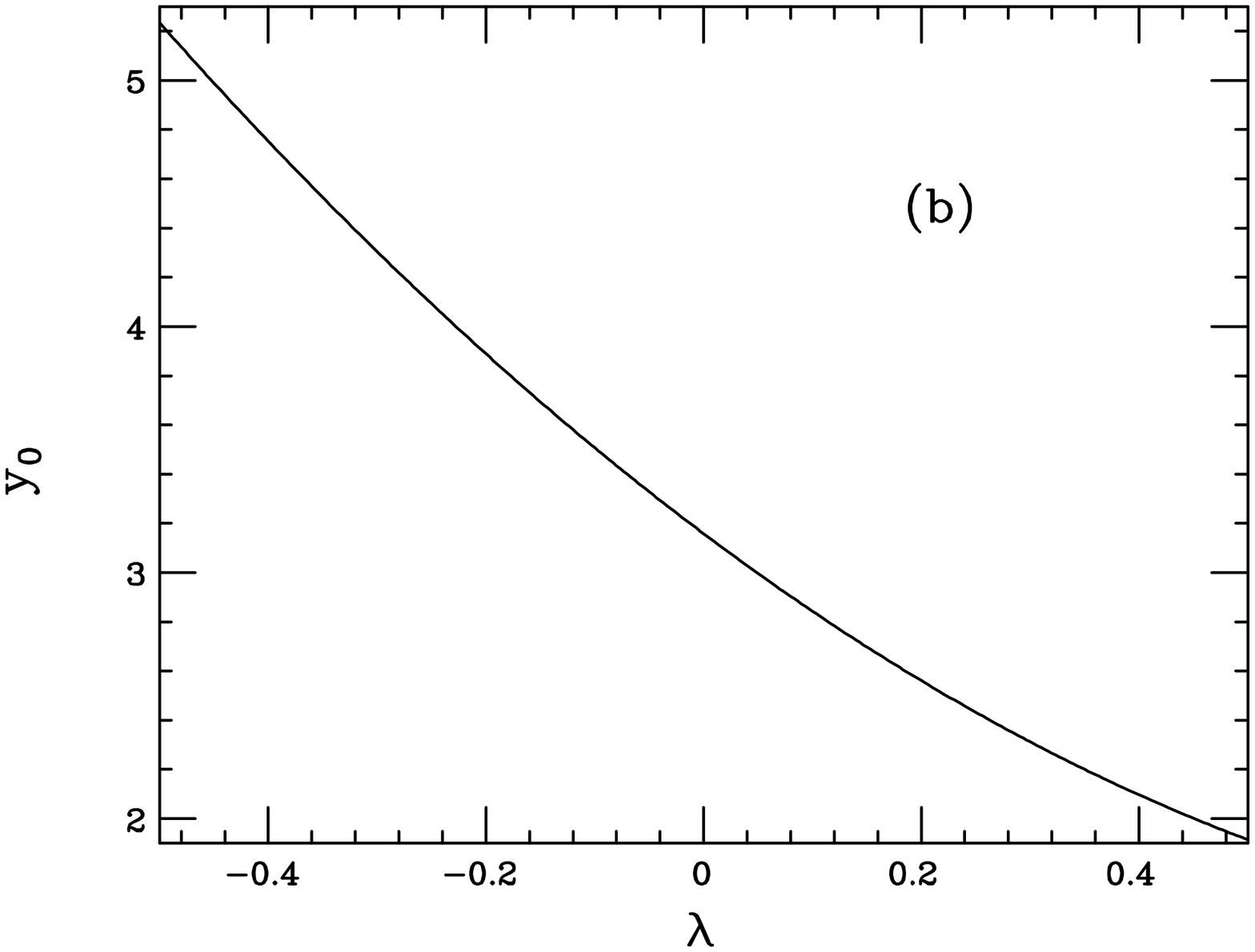,height=9.1cm,width=9.1cm,angle=90}}
\vspace*{-1cm}
\caption{Separate $\dk$ and $\lambda$ dependence of the
value of $y_0$, the zero position for the process $\gamma e \to W\nu$.}
\end{figure}
\vspace*{0.4mm}

One reason to believe {\it a priori} that $A_{pol}$, or $\sigma_{pol}$ itself,  
might be sensitive to modifications in the TGVs due to the presence of the 
anomalous couplings is the Drell-Hearn Gerasimov(DHG) Sum Rule{\cite {dhg}}.  
In its $\gamma e \to W \nu, Ze$ manifestation, the DHG sum rule implies that
\begin{equation}
\int_{1}^{\infty} {\sigma_{pol}(x)\over {x}} dx = 0 \,,
\end{equation}
for the tree level SM cross section when the couplings of all the 
particles involved in the process are `canonical', \ie, gauge invariant. 
That this integral is zero results from ($i$) the fact that 
$\sigma_{pol}$ is well 
behaved at large $x$ and ($ii$) a delicate cancellation occurs 
between the two 
regions where the integrand takes on opposite signs. This observation is 
directly correlated with the existence of a single, unique 
value of $x$(or $y$), \ie, $x_0$($y_0$),  where $\sigma_{pol}$(and, 
hence, $A_{pol}$) vanishes. For 
this reason $A_{pol}$ is sometimes referred to as $A_{DHG}$. 
For the $W(Z)$ case this asymmetry `zero' occurs at approximately 
$\sqrt {s_{\gamma e}}\simeq 254(150)$ GeV, both 
of which correspond to energies which are easily accessible at the NLC. In the 
$Z$ boson case the SM position of the zero can be obtained analytically as a 
function of the cut on the angle of the outgoing electron. In the 
corresponding $W$ case, 
the exact position of the zero can only be determined numerically.

\vspace*{-0.5cm}
\nn
\begin{figure}[htbp]
\centerline{
\psfig{figure=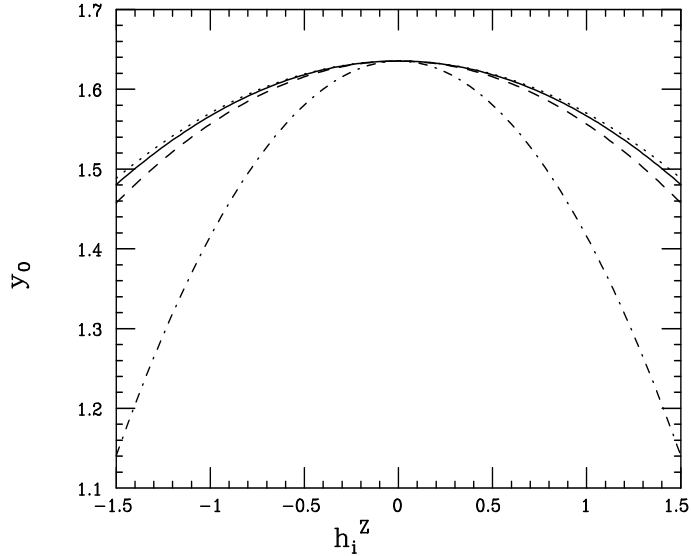,height=9.1cm,width=11cm,angle=-90}}
\vspace*{-1cm}
\caption{Position of the SM polarization asymmetry zero in 
$\gamma e \to Ze$ as a function of $h_{3,4}^0$ for $P=90\%$ with a $10^\circ$ 
angular cut. The dotted(dashed, 
dash-dotted, solid) curve corresponds to the case $h_4^0=0$($h_3^0=0$, 
$h_3^0=h_4^0$, $h_3^0=-h_4^0$).}
\end{figure}
\vspace*{0.4mm}

As discussed in detail in Ref.~{\cite {old}}, the inclusion of anomalous 
couplings not only moves the position of the zero but also forces the 
integral to become non-vanishing and, in most cases, 
{\it logarithmically divergent}. In fact, 
the integral is only finite when $\Delta \kappa+\lambda=0$, the same condition 
necessary for the existence of the radiation amplitude zero{\cite {raz}}. The 
reason for the divergence stems from the fact that the most divergent terms 
in $\sigma_{pol}$ proportional to the anomalous couplings become constants in 
the large $x$ limit; see Ref.~{\cite {old}} for complete expressions. It is 
interesting that the anomalous couplings do not induce additional zeros or 
extinguish the zero completely. 
Unfortunately, since we cannot go to infinite energies we cannot test the DHG 
Sum Rule directly but we {\it are} left with the position of the zero, or more 
generally, the asymmetry itself as a probe of TGVs. In the 
$W$ case the zero position, $y_0$, is found to be far more sensitive to 
modifications in the TGVs than is the zero position in in the $Z$ case. The 
zero position as a 
function of $\Delta \kappa$ and $\lambda$ for the $\gamma e\to W\nu$ process 
is shown in Fig.1 whereas the corresponding $Z$ case is shown in Fig.2. In 
either situation, the position of the zero 
{\it alone} does not offer sufficient sensitivity to the existence of anomalous 
couplings for us to obtain useful constraints.(See Ref. {\cite {old}.) 

\nn
\vspace*{0.1mm}
\hspace*{-0.5cm}
\begin{figure}[htbp]
\centerline{\psfig{figure=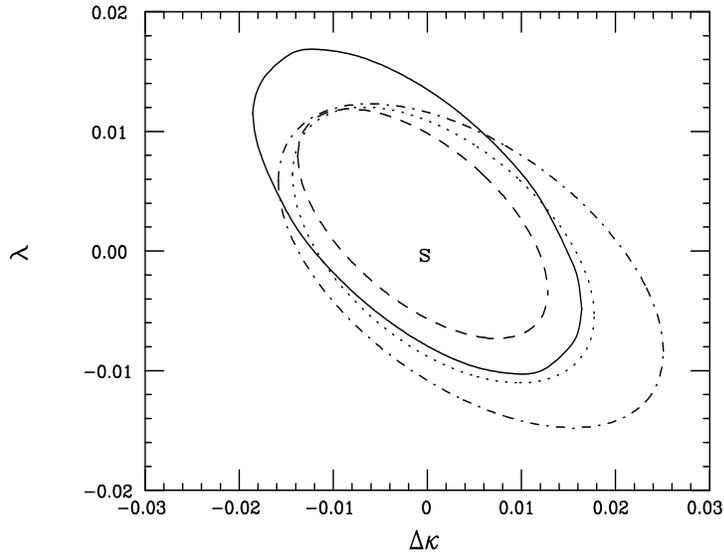,height=9.1cm,width=11cm,angle=90}}
\vspace*{-1.0cm}
\caption{95 $\%$ CL bounds on the $W$ anomalous couplings from the 
polarization asymmetry. The 
solid(dashed, dash-dotted) curves are for a 500 GeV NLC 
assuming complete $y$ coverage using 22(22, 44) bins and an integrated 
luminosity per bin of 2.5(5, 1.25)$fb^{-1}$, respectively. The corresponding 
bins widths are $\Delta y=$0.2(0.2, 0.1). The dotted curve 
corresponds to a 1 TeV NLC using 47 $\Delta y=0.2$ bins with 2.5 $fb^{-1}$/bin. 
`s' labels the SM prediction.}
\end{figure}

Our analysis begins by examining the energy, \ie, $y$ dependence of $A_{pol}$ 
for the two processes of interest; we consider the $W$ case first. For a 
500(1000) GeV collider, we see that only the range $1\leq y\leq 5.4(10.4)$ is 
kinematically accessible since the laser photon energy 
maximum is $\simeq 0.84E_e$. Since we are interested in bounds on the 
anomalous couplings, we will assume that the SM is valid and generate a set 
of binned $A_{pol}$ data samples via Monte Carlo taking 
only the statistical errors into account. We further assume that 
the electrons are 
90$\%$ left-handed polarized as right-handed electrons do not interact 
through the $W$ charged current couplings. Our bin width will be assumed to be 
$\Delta y=$0.1 or 0.2. 
We then fit the resulting distribution to 
the $\Delta \kappa$- and $\lambda$-dependent functional form of $A_{pol}(y)$ 
and subsequently 
extract the 95$\%$ CL allowed ranges for the anomalous couplings. The results 
of this procedure are shown in Fig. 3, where we see that reasonable 
constraints are obtained although only a single observable has been used in 
the fit. 

\vspace*{-0.5cm}
\nn
\begin{figure}[htbp]
\centerline{
\psfig{figure=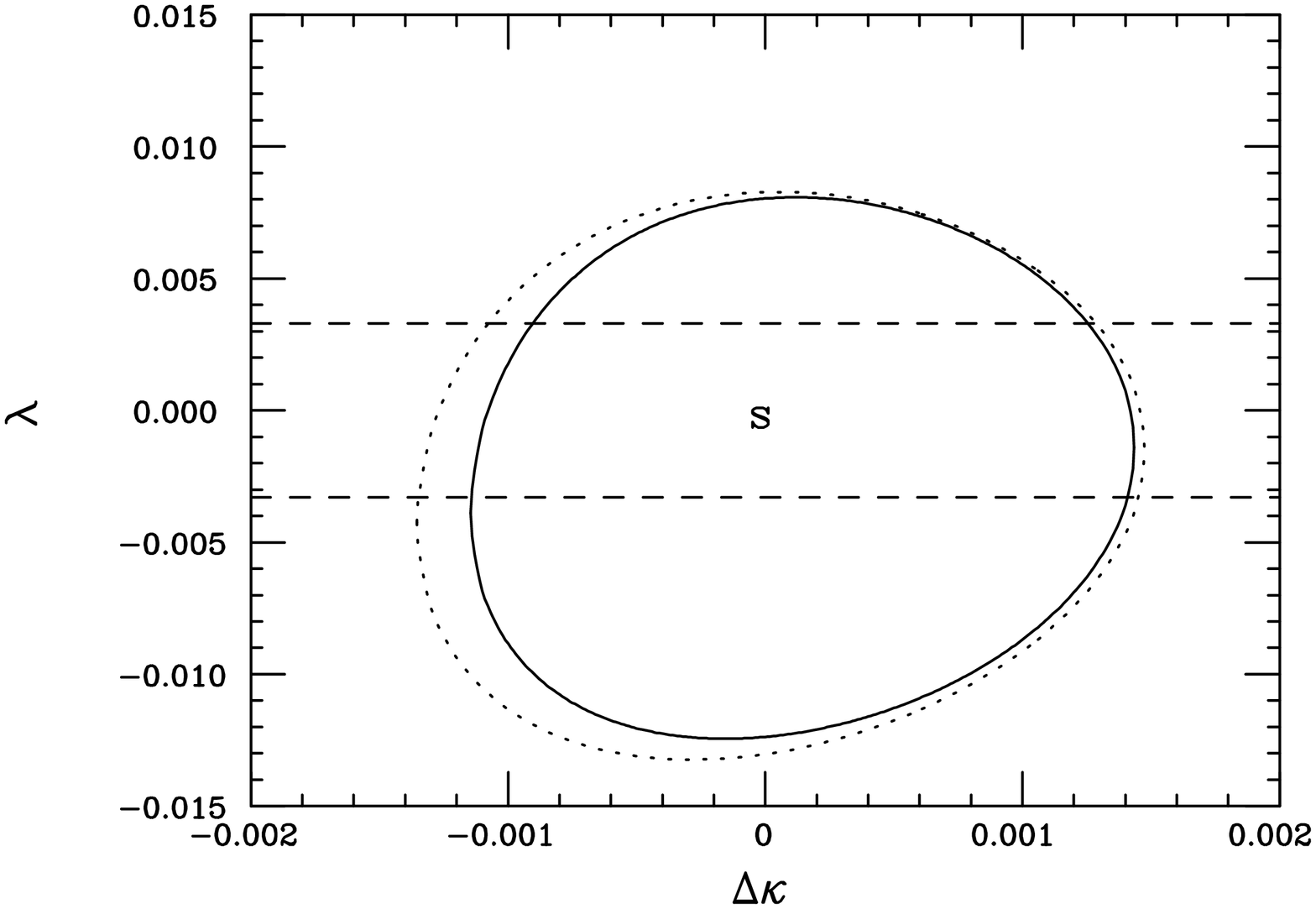,height=9.1cm,width=9.1cm,angle=-90}
\hspace*{-5mm}
\psfig{figure=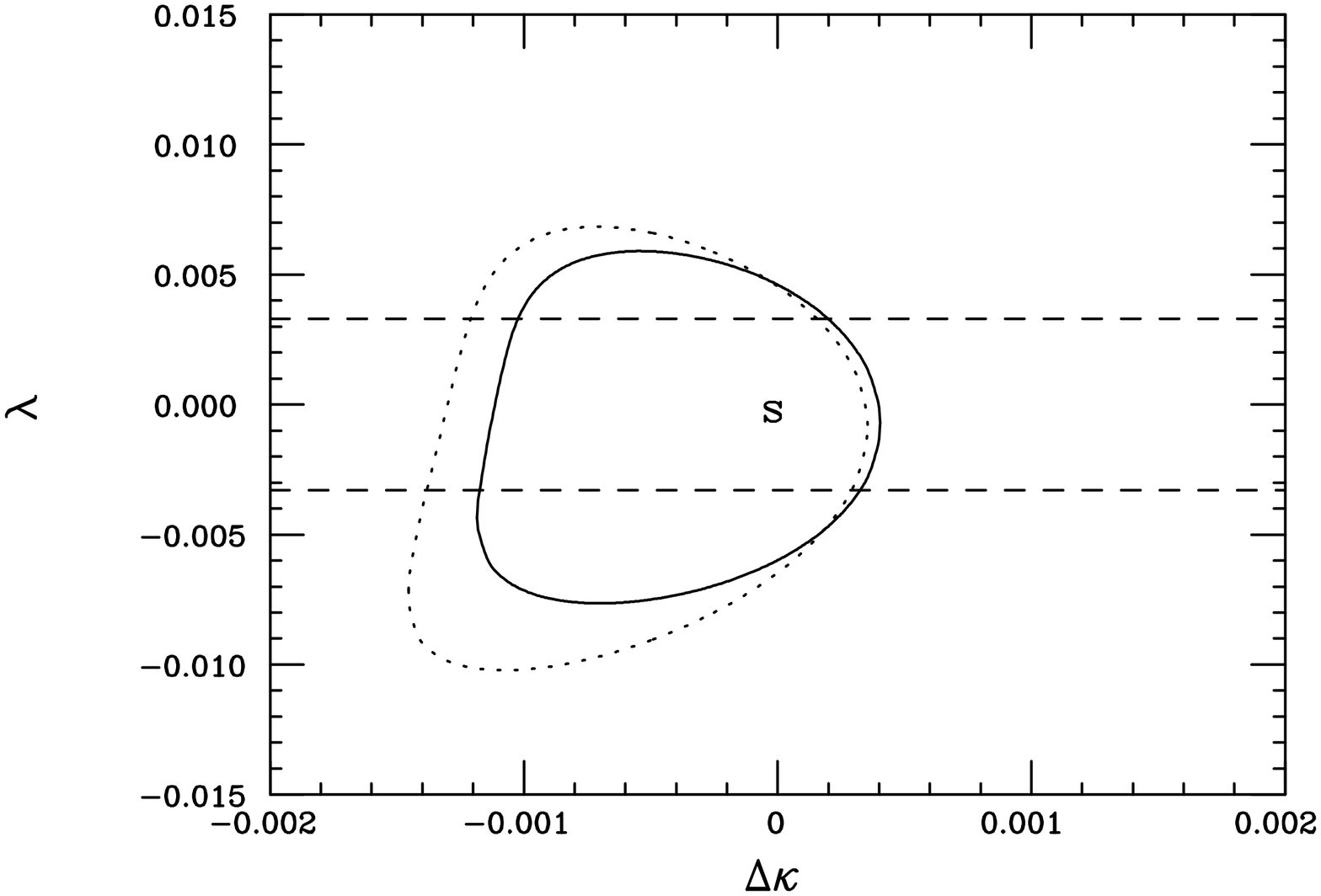,height=9.1cm,width=9.1cm,angle=-90}}
\vspace*{-1cm}
\caption{Same as the previous figure, but now for a (0.5)1 TeV NLC on the 
left(right) and combined with data 
on the total cross section and angular distribution in a simultaneous fit. The 
dotted(solid) curve uses the polarization 
asymmetry and total cross section(all) data. Only 
statistical errors are included. The dashed lines are the corresponding 
bounds from the LHC from the $pp\to W\gamma +X$ process with an integrated 
luminosity of 100 $fb^{-1}$.}
\end{figure}
\vspace*{0.4mm}

Clearly, to obtain stronger limits we need to make a combined fit with other 
observables, such as the energy dependence of the total cross section, the 
$W$ angular distribution, or the net $W$ polarization. As an example we 
show in Fig. 4 from the results of our Monte Carlo study that the size of 
the $95\%$ CL allowed region shrinks drastically in the both the 0.5 and 
1 TeV cases when the $W$ angular distribution and energy-dependent total 
cross section data are included in a simultaneous fit together 
with the polarization asymmetry. For this analysis the angular distribution 
was placed into 10 bins and energy averaged over the accessible kinematic 
region. The total cross section data was binned in exactly the same way as 
was the polarization asymmetry. Note that the constraints obtained by this 
analysis are superior to that of the LHC{\cite {rev}} with an integrated 
luminosity of 
100 $fb^{-1}$. (The LHC constraints on $\Delta \kappa$ are rather poor whereas 
the $\lambda$ bounds are somewhat better.)
As is well known, both the total cross section and the $W$ angular 
distributions are highly sensitive to 
$\Delta \kappa$ and thus the allowed region is highly compressed in that 
direction. At 500 GeV(1 TeV), we find that $\Delta \kappa$ is bounded 
to the range 
$-1.2\cdot 10^{-3}\leq \Delta \kappa \leq 1.4(0.4)\cdot 10^{-3}$ while the 
allowed $\lambda$ range is still rather large. Further improvements in these 
limits will result from data taken at a 1.5 TeV NLC.

\vspace*{-0.5cm}
\nn
\begin{figure}[htbp]
\centerline{
\psfig{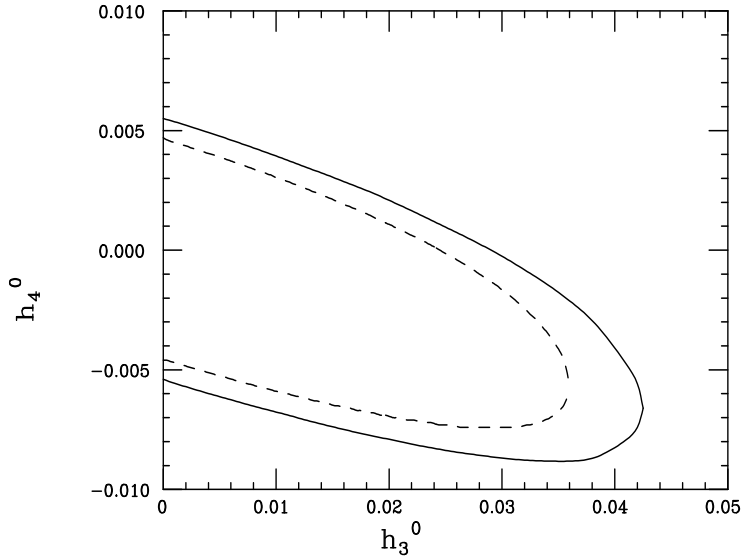}}
\vspace*{-1cm}
\caption{$95\%$CL allowed region for the anomalous coupling parameters 
$h_3^0$ and $h_4^0$ from a combined fit to the energy dependencies of the total 
cross section and polarization asymmetry at a 500 GeV NLC assuming $P=90\%$ 
and an integrated luminosity of $3(6)fb^{-1}$/bin corresponding to the solid
(dashed) curve. 18 bins of width $\Delta y$=0.2 were chosen to cover the $y$ 
range $1\leq y \leq 4.6$. The corresponding bounds for negative values of 
$h_3^0$ are obtainable by remembering the invariance of the polarization 
dependent cross section under the reflection $h_{3,4}^0\to -h_{3,4}^0$.}
\end{figure}
\vspace*{0.4mm}

With this experience in mind, in the $Z$ case we will follow a similar approach 
but we 
will simultaneously fit both the energy dependence of $A_{pol}$ as well as 
that of 
the total cross section. (Later, we will also include the $Z$ boson's angular 
distribution 
into the fit.) In this $Z$ analysis we make a $10^{\circ}$ angular cut on the 
outgoing electron and keep a finite form factor scale, $\Lambda=1.5$ TeV, so 
that we may more readily compare with other existing analyses. (The angular 
cut also gives us a finite cross section in the massless electron limit; 
this cut is not required in the case of the $W$ production process.) We again  
assume that $P=90\%$ so that data taking for this analysis can take place 
simultaneously with that for the $W$. The accessible $y$ ranges are now 
$1\leq y \leq 4.6(9.4)$ for a 500(1000) GeV collider. Fig.5 shows our results 
for the 500 GeV NLC while Fig.6 shows the corresponding 1 TeV case. For a 
given energy and fixed total integrated luminosity we learn from these figures 
that it is best to 
take as much data as possible at the highest possible values of $y$. 
Generally, one finds that increased sensitivity to the existence of anomalous 
couplings occurs at the highest possible collision energies.

\vspace*{-0.5cm}
\nn
\begin{figure}[htbp]
\centerline{
\psfig{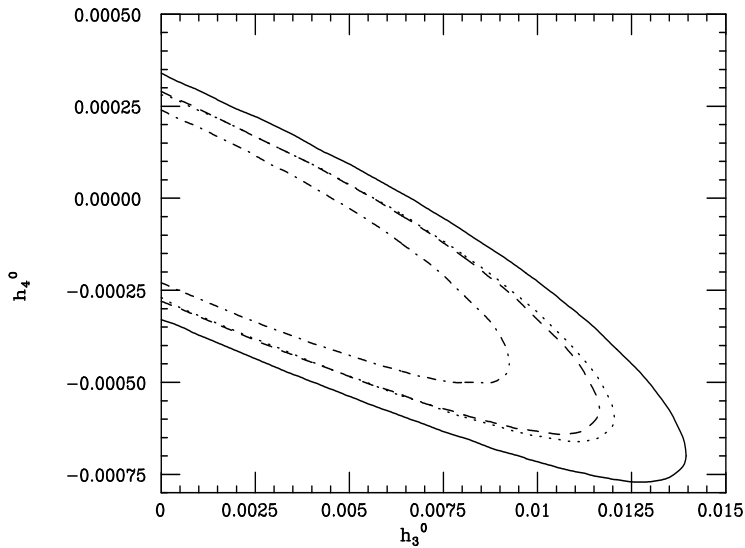}}
\vspace*{-1cm}
\caption{Same as Fig. 5 but for a 1 TeV NLC. The solid(dashed) curve 
corresponds to a luminosity of $4(8)fb^{-1}$/bin for 42 bins of width 
$\Delta y$=0.2 which covered the range $1\leq y \leq 9.4$. The dotted curve 
corresponds to a luminosity of $8fb^{-1}$/bin but only for the last 21 bins. 
The dash-dotted curve corresponds to the case of $16.8fb^{-1}$/bin in only the 
last 10 bins.}
\end{figure}
\vspace*{0.4mm}

Even these anomalous coupling bounds can be significantly 
improved by including the $Z$ boson angular information in 
the fit. To be concrete we examine the case of a 1 TeV NLC with 
16.8$fb^{-1}$/bin of integrated luminosity taken in the last 10 $\Delta y$ 
bins(corresponding to the dash-dotted curve in Fig.6). Deconvoluting the 
angular integration and performing instead the integration over the 10 
$\Delta y$ bins we obtain the energy-averaged angular distribution. Placing 
this distribution into 10 (almost) equal sized $cos \theta$ bins while still 
employing our $10^\circ$ cut, we can use this 
additional data in performing our overall simultaneous $\chi^2$ fit. The 
result of this procedure is shown in Fig.7 together with 
the anticipated result from the LHC using the $Z\gamma$ production mode. Note 
that the additional angular distribution data has reduced the size of the 
$95\%$ CL allowed region by almost a factor of two. 
Clearly both machines are complementary in their abilities to probe small 
values of the $\gamma ZZ$ anomalous couplings. As in the $W$ case, if 
the NLC and LHC results were 
to be combined, an exceptionally small allowed region would remain. 
The NLC results themselves may be further improved by considering 
measurements of the polarization of the final state $Z$ as well as 
by an examination of, \eg, the complementary $e^+e^- \to Z\gamma$ process.

\vspace*{-0.5cm}
\nn
\begin{figure}[htbp]
\centerline{
\psfig{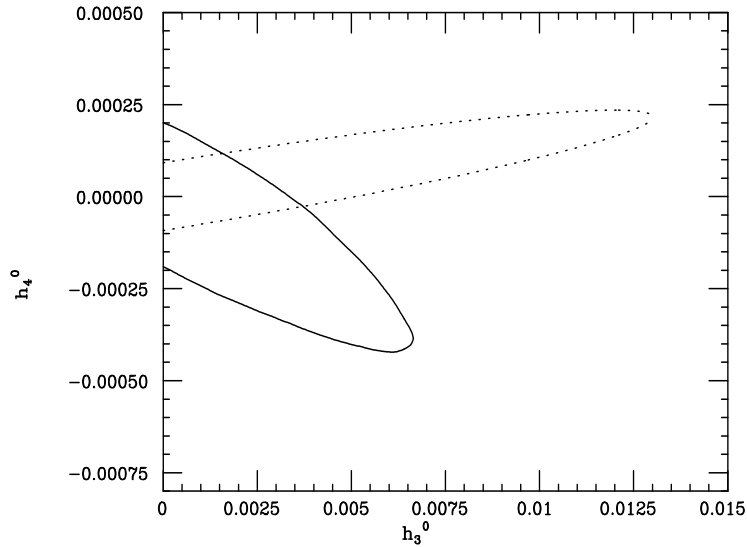}}
\vspace*{-1cm}
\caption{The solid curve is the same as dash-dotted curve 
in Fig. 6, but now including in the fit 
the $Z$ boson angular distribution obtained from the highest 
10 bins in energy. The 
corresponding result for the 14 TeV LHC with 100$fb^{-1}$ of integrated 
luminosity from the process $pp\to Z\gamma+X$ is shown as the dotted curve.}
\end{figure}
\vspace*{0.4mm}

\section{Discussion and Conclusions}

The collision of polarized electron and photon beams at the NLC offers an 
exciting opportunity to probe for anomalous gauge couplings of both the $W$ 
and the $Z$ through the use 
of the polarization asymmetry. In the case of $\gamma e \to W\nu$ we can 
cleanly isolate the $\gamma WW$ vertex in a model independent fashion. When 
combined with other observables, extraordinary sensitivities to such 
couplings for $W$'s are achievable at the NLC in the $\gamma e$ mode. These are 
found to be quite complementary to those obtainable in $e^+e^-$ collisions as 
well as at the LHC. In the case of the 
$\gamma ZZ$ anomalous couplings, we found constraints comparable to those 
which can be obtained at the LHC.

\Acknowledgements

The author would like to thank S. J. Brodsky, I. Schmidt, J.L. Hewett, and   
S. Godfrey for discussions related to this work.

%
\def\MPL #1 #2 #3 {Mod.~Phys.~Lett.~{\bf#1},\ #2 (#3)}
\def\NPB #1 #2 #3 {Nucl.~Phys.~{\bf#1},\ #2 (#3)}
\def\PLB #1 #2 #3 {Phys.~Lett.~{\bf#1},\ #2 (#3)}
\def\PR #1 #2 #3 {Phys.~Rep.~{\bf#1},\ #2 (#3)}
\def\PRD #1 #2 #3 {Phys.~Rev.~{\bf#1},\ #2 (#3)}
\def\PRL #1 #2 #3 {Phys.~Rev.~Lett.~{\bf#1},\ #2 (#3)}
\def\RMP #1 #2 #3 {Rev.~Mod.~Phys.~{\bf#1},\ #2 (#3)}
\def\ZP #1 #2 #3 {Z.~Phys.~{\bf#1},\ #2 (#3)}
\def\IJMP #1 #2 #3 {Int.~J.~Mod.~Phys.~{\bf#1},\ #2 (#3)}

\end{document}